\documentclass[12pt]{article}
\usepackage{amssymb}
\usepackage{amsmath}
\usepackage{graphicx}
\usepackage{graphics}
\usepackage{tikz}
\usepackage{float}
\begin{document}
\title{\textbf{Reductive Perturbation Method in Magnetized Plasma and Role of Negative Ions}}
\author{H. Saleem$^{1,2,3}$, Shaukat Ali Shan$^{1,4}$, and S. Poedts$^{5,6}$ \\
%EndAName
$^{1}$Theoretical Research Institute, Pakistan Academy of Sciences, \\
3-Constitution Avenue, G-5/2, Islamabad (44000), Pakistan,\\
$^{2}$Department of Physics, School of Natural Sciences (SNS), \\
National University of Science and Technology (NUST), \\
Islamabad (44000), Pakistan,\\
$^{3}$Space and Astrophysics Research Lab. (SARL), \\National Centre of GIS and Space Applications, \\Islamabad (44000), Pakistan,\\
$^{4}$Theoretical Physics Division (TPD), PINSTECH, \\
P. O. Nilore, Islamabad (45650), Pakistan,\\
$^{5}$Centre for mathematical Plasma Astrophysics, KU Leuven, \\
Celestijnenlaan 200b, 3001 Leuven, Belgium,\\
$^{6}$Institute of Physics, University of Maria Curie-Sk{\l}odowska, \\ 
ul.\ Radziszewskiego 10, 20-031 Lublin, Poland\\
\textbf{Email:} saleemhpk@hotmail.com; \\ shaukatshan@gmail.com; Stefaan.Poedts@kuleuven.be}
\maketitle

\begin{abstract}
An analysis of reductive perturbation method (RPM) is presented to show that why the solitary structures of nonlinear ion acoustic waves (IAWs) cannot be obtained in magnetized electron ion plasma by employing this technique. In RPM, the nonlinear Korteweg-de Vries (KdV) equation is derived using stretched coordinates in the reference frame of the wave phase speed, considering the dispersion to be a higher-order effect which balances the nonlinearity to produce a solitary structure. The maximum amplitude $\mid \Phi_m \mid$ of the nonlinear solitary wave turns out to be larger than one which contradicts the small amplitude approximation. In the presence of negative ions, the maximum amplitude satisfies the condition $\mid\Phi_m\mid <1$.
To elaborate these points, the results have been applied to an experimental plasma consisting of positive ions of xenon $(Xe^{+})$ and negative ions of fluorene $(F^{-})$ along with electrons. The amplitude and width of the solitary structures depend upon the ratio of the electron to positive ion density ($\frac{n_{e0}}{n_{i0}}$). Since the nonlinear coefficient turns out to be negative, rarefied (dip) solitons are formed in the magnetized $Xe^{+}-F^{-}-e$ plasma.
\end{abstract}

%%%%%%%%%%%%%%%%%%%%%%%%%%%%%%%%%%%%%%%%%%%%%%%%%%%%%%%%%%%%%%%%%%%%%%%
\section{Introduction}
%%%%%%%%%%%%%%%%%%%%%%%%%%%%%%%%%%%%%%%%%%%%%%%%%%%%%%%%%%%%%%%%%%%%%%%
The reductive perturbation method (RPM) was employed to investigate the formation of 
electrostatic solitary structures by the nonlinear ion acoustic waves (IAWs)
in the small amplitude limit in unmagnetized $({\bf B}_0=0)$ plasma long ago \cite{Washimi1966}.
Recently, the Korteweg-de Vries (KdV) and modified Korteweg-de Vries (m-KdV) equations have been obtained for IAWs in unmagnetized plasma having negative ions as well \cite{Madhukalya2023}. Solitary structures of IAWs in magnetized electron positron ion plasma have also been investigated \cite{Gill2009} assuming ions to be relativistic with non-zero positron density. But 
in classical magnetized (${\bf B}_0 \neq 0$) electron ion plasma, the solitary structures of IAWs were studied using 
the Sagdeev potential approach, i.e.\ assuming arbitrary amplitude of the nonlinear waves 
\cite{Sagdeev1966, Sultana2010}. The Korteweg-de Vries-Burgers (KdVB) equation has also been obtained for IAWs in magnetized electron ion plasma using the RPM and the resulting shock waves have been investigated \cite{Shahmansouri2013}. But the KdV solitons have not been discussed in the limit of dissipation-less plasma. We have noticed that there appears an inconsistency in the RPM method when it is employed to magnetized plasma in the small amplitude limit. The nonlinear wave dynamics in the framework of RPM is investigated in the frame of the phase speed of the wave which in the lowest order does not contain the contribution of the ions polarization drift in case of magnetized electron ion plasma. The nonlinear coefficient in KdV equation turns out to be smaller than one because wave propagates obliquely making an angle with the ambient magnetic field ${\bf B}_0$. Consequently, the maximum normalized amplitude  of the solitary pulse obtained by Korteweg-de Vries (KdV) equation becomes greater than one which is a contradiction to the small amplitude limit. This fact has not been pointed out in literature so far to the best of the author's knowledge.
This seems to be the reason that solitary structures of IAWs are not obtained in classical magnetized electron ion plasma using the RPM.

On the other hand, the characteristics of waves are modified drastically when negative ions are introduced in the plasma \cite{Sal2022}.
Plasmas with negative and positive ions (NPI) along with electrons were generated long ago in the Q-machine in the USA at Princeton \cite{Goe1966, Dan1966}. After several years, another experimental study of NPI plasma was performed \cite{Bac1979}. The aim of these experiments and
investigations was to study the effects of the presence of negative ions on the plasma
dynamics. Later on, series of experiments were performed to produce pure NPI
plasmas in Iowa \cite{Son1991, Kim2007, Kim2008, Kim2013}. Negative
ions are commonly observed in space and astrophysical plasmas including plasmas in the
terrestrial ionosphere, dusty plasmas of planetary magnetospheres and
interstellar medium (ISM) \cite{Nar1971, Mas1976, Rei1990}. The charged
particles system containing negative and positive ions along with electrons is generally called the negative-positive-ion-electron (NPIE) plasma. On the other hand,
high energy gamma radiation in strongly magnetized astrophysical
environments such as accretion disks, active galactic nuclei (AGN), and
magnetospheres of neutron stars produce electron positron (EP) pair plasmas.
The characteristics of EP plasma have been analyzed along with the study of
linear and nonlinear wave dynamics by several authors \cite{Iwa1993,
Shu1986, Sto2020}. Electron positron (EP) plasmas were also created in
laboratories at low densities, $n_0\simeq 10^{8}\;$cm$^{-3}$ \cite{Sur1990,
Gre1997}. However, the EP plasma confinement at higher densities and for
longer periods of time is difficult due to $\gamma$-ray annihilation
problem. Therefore, a group of scientists in Japan tried to produce pure
pair ion (PI) plasma having equal mass negative positive ions $(m_{+}=m_{-})$
because such plasmas can be confined for longer periods of time and possibly
the properties of pair plasmas can be investigated in detail. Several
experiments have been performed to create pure PI plasmas of fullerenes $%
C_{60}^{\pm}$ \cite{Ooh2003, Ooh2005} as well as of hydrogen $H^{\pm}$ and
helium $He^{\pm}$ \cite{Ooh2007, Ooh2019}. Longitudinal waves were excited
in PI plasma of fullerenes and ion acoustic wave was also observed \cite%
{Ooh2005}.

A few authors pointed out \cite{SVP2006} that the observation of
IA waves in the experiment \cite{Ooh2005} was an indication that the produced 
fullerene plasma was not a pure PI plasma, rather electrons were also present 
in significant proportion in the system. 
However, these authors
used the quasi-neutrality approximation for investigating ion acoustic waves in
the pair ion plasma including the effects of electrons. Dispersion
relations of a few other low frequency modes were also discussed. Later on,
one of the authors \cite{Sal2006} pointed out that quasi-neutrality was 
not a good
approximation to investigate the waves in PIE plasma. The arguementation was that when
the electron density $n_{e0}$ reduces significantly in the PI plasma,
the electron Debye length $\lambda_{De}=(\frac{T_{e0}}{4 \pi n_{e0} e^2})^{1/2}$
becomes very large and electrons cannot be responsible for the shielding. 

In 2007, a quantitative criterion was presented to define pure PI plasma 
\cite{Sal2007} and it was pointed out that the ratio of electron density to
positive ion density, $(n_{e0}/n_{+0})$, is crucial to decide whether the
produced plasma in laboratory can be considered as a pure PI plasma or not. The author
pointed out that the electron dynamics in the plasma can be ignored only if the
condition $\omega _{pe}\ll\omega _{p\pm }$ holds where $\omega _{pe}=(\frac{4
\pi n_{e0}e^2}{me})^{1/2}$ is electron plasma oscillation frequency while $%
\omega _{p\pm }=(\frac{4 \pi n_{\pm}e^2}{m_{\pm}})^{1/2}$ denote the
positive and negative ion plasma oscillation frequencies, respectively. This
condition requires $n_{e0}/n_{\pm 0}\ll m_{e}/m_{\pm }$, where $m_{e}$ is
electron mass, $m_{+}$ denotes the positive ion mass and $m_{-}$ corresponds to the negative ion
mass. Since $n_{0+}=n_{e0}+n_{-0}$, the simplest form of the
criterion for pure PI plasma can be expressed as $\omega _{pe}\ll \omega _{p+}$
. This criterion is also valid for the negative positive ion plasma. Since
the electron mass is very small compared to the proton mass, it is very
difficult to achieve this criterion in the laboratory. The role of the electron temperature and its effects on IA waves in addition to the density ratio $%
(n_{e0}/n_{+0})$, have also been discussed in that paper. In a
previous investigation of the same author, it was pointed out by using kinetic
theory that the Landau damping rate of ion acoustic wave decreases when negative ions are present and, hence, this wave can be excited in
such plasma systems easily \cite{Sal2006}. Low frequency electrostatic
drift waves \cite{Ali2010, Ros2013} and the effects of field-aligned shear flow on
ion acoustic wave instability in PIE and NPIE plasmas have been
investigated using the kinetic approach \cite{Sha2019}. Nonlinear structures, such as solitons
and vortices, were also explored in PIE plasma \cite{Sal2007, Shu2005, Sal2017}. A few authors \cite{Sch2005, Kon2014} tried to explain the observations from the experiment \cite{Ooh2005} using kinetic theory, assuming that the generated plasma with fullerene ions was pure PI plasma.

In an experiment with hydrogen PI plasma, the electron density has been
estimated to be about hundred times lower than the positive ion density $%
n_{e0}/n_{+0}<10^{-2}$ \cite{Ooh2019}, where $n_{e0}$ and $n_{+0}$ are the
equilibrium electron and positive ion densities, respectively. However, these authors pointed out that the criterion presented in Ref.~\cite{Sal2007} for a pure PI plasma was not achieved.
On the other
hand, in the experiment on NPI plasma with positive potassium ions $K^{+}$
and negative ions of perfluoromethylcyclohexane $C_{7}F_{14}^{-1}$, the
authors estimated the ratio of electron to positive ion density to be $%
n_{e0}/n_{+0}\leq 10^{-4}$ \cite{Kim2013}, with the aim to produce pure NPI plasma. Even if the electron densities in the
above mentioned plasmas are much smaller than the positive ion densities ($n_{e0}\ll n_{+0}$), 
these plasmas
cannot be defined as pure ionic plasmas. The reason is explained below.

Since potassium ion has mass $m_{+}\simeq 39m_{p}$ and $C_{7}F_{14}^{-1}$
has $m_{-}\simeq 350m_{p}$, where $m_{p}$ denotes the mass of a proton, therefore $%
m_{e}/m_{+}\simeq (1.39)\times 10^{-5}$ and $m_{e}/m_{-}\simeq (1.55)\times
10^{-6}$. This implies that $\omega _{p\pm }\ll \omega _{pe}$ in these
experiments, and hence, even if $\frac{n_{e0}}{n_{i0}}\ll 1$ holds they cannot be defined as pure ionic plasmas. In several experimental \cite{
Kim2008, Kim2013, Ooh2005, Ooh2007} and theoretical \cite{Shu2005, Sch2005,
Kon2014} research papers including a widely used text book \cite{Che2013},
the wave analysis of PI and NPI plasmas has been presented ignoring the criterion for ionic plasmas. Recently \cite%
{Sal2022}, the possible ion modes in NPIE and PIE plasmas have been
discussed in detail. Limiting cases of pure NPI and pure PI plasmas have also
been obtained. A large number of research papers on
wave dynamics in NPIE and PIE plasmas have appeared in the literature \cite%
{Vra2005, Ver2006, Kou2006, Kou2007, Mah2008, Vra2008a, Vra2008b}.

It is interesting to note that the KdV equation derived for IAWs in usual 
magnetized electron ion
plasma under RPM, does not yield a soliton structure in the simplest case of Boltzmann electrons.  Possible explanations are given in sections 4 and 5. The IAWs were observed in NPIE plasma produced in Japan with positive ions of xenon $Xe^{+}$ and negative ions of fluorene $F^{-}$ and electrons \cite{Ichiki2001, Ichiki2002}. These authors focused their studies on the characteristics of linear IAWs in unmagnetized NPIE plasma and did not discuss the pure NPI plasmas limit and nonlinear propagation. They also produced NPIE plasma with different atoms and molecules.
 
In the next section, the theoretical model along with the basic set of normalized equations is presented. In section~3, the KdV equation for small amplitude ion acoustic waves in magnetized
negative positive ion electron (NPIE) plasma is derived using the reductive perturbation method considering
the electrons to be inertia-less. Some important comments on the small amplitude ion acoustic waves in magnetized $Xe^{+}-F^{-}-e$ plasma are highlighted in section~4. The numerical results are presented in
section~5. Finally, the results are discussed both qualitatively and
quantitatively in section~6.

%%%%%%%%%%%%%%%%%%%%%%%%%%%%%%%%%%%%%%%%%%%%%%%%%%%%%%%%%%%%%%%%%%%%%%%
\section{Theoretical Model}
%%%%%%%%%%%%%%%%%%%%%%%%%%%%%%%%%%%%%%%%%%%%%%%%%%%%%%%%%%%%%%%%%%%%%%%
Let us consider negative positive ion electron (NPIE) plasma
immersed in a constant external magnetic field ${\bf B}_{0}=B_{0}\hat{z}$, where $\hat{z}$ is a unit vector along $z$-axis. The nonlinear
dynamics of the low frequency purely electrostatic perturbations can be
described by the following set of normalized equations,

\begin{equation}
\frac{\partial n_{i}}{\partial t}+\nabla \cdot (n_{i}\mathbf{v}_{i})=0,
\label{1}
\end{equation}

\begin{equation}
\frac{\partial \mathbf{v}_{i}}{\partial t}+(\mathbf{v}_{i}\cdot \nabla )%
\mathbf{v}_{i}=-\mathbf{\nabla }\Phi +\sigma _{i}(\mathbf{v}_{i}\times \hat{z%
})-2\theta _{i}\mathbf{\nabla }n_{i},  \label{2}
\end{equation}

\begin{equation}
\frac{\partial n_{n}}{\partial t}+\nabla \cdot (n_{n}\mathbf{v}_{n})=0,
\label{3}
\end{equation}

\begin{equation}
\frac{\partial \mathbf{v}_{n}}{\partial t}+(\mathbf{v}_{n}\cdot \nabla )%
\mathbf{v}_{n}=\alpha _{pn}\mathbf{\nabla }\Phi -\sigma _{n}(\mathbf{v}%
_{n}\times \hat{z})-2\alpha _{pn}\theta _{n}\mathbf{\nabla }n_{n}.  \label{4}
\end{equation}

The lighter electrons are assumed to follow the Maxwell-Boltzmann relation,

\begin{equation}
n_{e}=n_{e0}e^{\Phi }.  \label{5}
\end{equation}
In this case, the Poisson equation can be written as, 
\begin{equation}
\nabla \cdot \mathbf{E}=n_{i}-N_{e0}n_{e}+N_{n0}n_{n},  \label{6}
\end{equation}
where subscripts $i, n, e$ denote positive ions, negative ions and electrons,
respectively. Furthermore, $n_{e0}/n_{i0}=N_{e0}$, $n_{n0}/n_{i0}=N_{n0}$, $\alpha _{pn}=m_{i}/m_{n}$, $\sigma _{j}=\Omega _{j}/\omega _{pj}$, $%
\Omega _{j}=eB_{0}/cm_{j}$, $\theta _{j}=T_{j}/T_{e}$, $\gamma _{j}=(N+2)/N$  
where $N$ is the number of degrees of freedom and $j=i, n$. We consider wave propagation in the $yz$-plane i.e.\ $\nabla =(0, \partial_y, \partial_z)$. Therefore,  $N=2$  and, hence, $\gamma _{j}=2$. The quantities $\mathbf{%
v}_{i}$ and $\mathbf{v}_{n}$ are the positive and negative ion fluid speeds
normalized by $c_{si}=\sqrt{T_{e}/m_{i}}$, $\Phi $ is the electrostatic wave
potential normalized by $T_{e}/e$, while the time variable $t$ is normalized by $%
t_{pi}=\sqrt{m_{i}/4\pi n_{i0}e^{2}}=(\omega_{pi})^{-1}$, and the space variable $r$ is normalized
by $\lambda _{D}=c_{si}/\omega _{pi}$.

%%%%%%%%%%%%%%%%%%%%%%%%%%%%%%%%%%%%%%%%%%%%%%%%%%%%%%%%%%%%%%%%%%%%%%%
\section{Derivation of KdV equation for IAW in NPIE Plasma}
%%%%%%%%%%%%%%%%%%%%%%%%%%%%%%%%%%%%%%%%%%%%%%%%%%%%%%%%%%%%%%%%%%%%%%%

In order to derive the nonlinear dynamical Korteweg-de Vries equation in
magnetized NPIE plasma, we adopt the standard reductive perturbation method (RPM).
The stretched coordinates are defined as, 
\begin{equation}
\xi =\epsilon ^{^{1/2}}(l_{y}y+l_{z}z-\lambda t)\text{, and }\tau =\epsilon
^{^{3/2}}t,  \label{8}
\end{equation}%
where $\epsilon $ is a small $(0<\epsilon < 1)$ expansion parameter
characterizing the strength of the nonlinearity and $\lambda $ is the phase
velocity of the wave normalized with acoustic speed corresponding to
positive ions $c_{si}$. The $l_{y}$ and $l_{z}$ are, respectively, the
direction cosines such that $l_{y}^{2}+l_{z}^{2}=1$.

Now, using RPM we can expand the
perturbed quantities about their equilibrium values in powers of $%
\epsilon $ as follows \cite{Shahmansouri2013}, 
\begin{align}
n_{j}& =1+\epsilon n_{j1}+\epsilon ^{2}n_{j2}+\ldots,  \notag \\
v_{jx}& =\epsilon ^{^{3/2}}v_{jx1}+\epsilon ^{2}v_{jx2}+\ldots,  \notag \\
v_{jy}& =\epsilon ^{^{3/2}}v_{jy1}+\epsilon ^{2}v_{jy2}+\ldots,  \notag \\
v_{jz}& =\epsilon v_{jz1}+\epsilon ^{2}v_{jz2}+\ldots,  \notag \\
\Phi & =\epsilon \Phi _{1}+\epsilon ^{2}\Phi _{2}+\epsilon ^{3}\Phi _{3}\ldots.
\label{9}
\end{align}%
The lowest order terms of the equations of motion and mass conservation $%
\epsilon^{3/2} $ lead to,

\begin{equation}
n_{i1}=\frac{l_{z}^{2}\Phi _{1}}{\lambda _{mp}^{2}}\text{, and }v_{iz1}=%
\frac{\lambda l_{z}\Phi _{1}}{\lambda _{mp}^{2}} , \label{10}
\end{equation}%
\begin{equation}
n_{n1}=-\frac{\alpha _{pn}l_{z}^{2}}{\lambda _{mn}^{2}}\Phi _{1}\text{, and }%
v_{nz1}=-\frac{\alpha _{pn}\lambda l_{z}}{\lambda _{mn}^{2}}\Phi _{1},
\label{11}
\end{equation}%
and%
\begin{equation}
n_{e1}=\Phi _{1},
\label{12}
\end{equation}%
where $\lambda _{mp}^{2}=(\lambda ^{2}-2\theta _{i}l_{z}^{2})$ and $\lambda _{mn}^{2}=(\lambda
^{2}-2\alpha _{pn}\theta _{n}l_{z}^{2})$. The lowest order terms in the Poisson equation
are the $\epsilon$-order terms, which give the quasi-neutrality under the above
mentioned ordering, 
\begin{equation}
N_{e0}n_{e1}+N_{n0}n_{n1}-n_{i1}=0.  \label{13}
\end{equation}%
The linear dispersion relation can be obtained by substituting the values of 
$n_{i1}$, $n_{n1}$ and $n_{e1}$ in Eq.~(12), 
\begin{equation}
-N_{e0}+\frac{N_{n0}\alpha
_{pn}l_{z}^{2}}{\lambda _{mn}^{2}}+\frac{l_{z}^{2}}{\lambda _{mp}^{2}}=0.
\label{14}
\end{equation}%
Note that the dispersive effects of the ions polarization drift as well as of the term $\nabla \cdot {\bf E} \neq 0$, do not appear in the above dispersion relation.

The lowest order terms of the momentum equations in the $x$ and $y$-components, are of the order $\epsilon ^{3/2}$, which yields,

\begin{equation}
v_{ix1}=-\frac{l_{y}}{\omega _{i}}\left[ 1+\frac{2 \theta
_{i}l_{z}^{2}}{\lambda _{mp}^{2}}\right] \frac{\partial \Phi _{1}}{\partial
\xi },  \label{15}
\end{equation}%
and 
\begin{equation}
v_{nx1}=-\frac{\alpha _{pn}l_{y}}{\omega _{n}}\left[ 1+\frac{\alpha
_{pn}2\theta _{n}l_{z}^{2}}{\lambda _{mn}^{2}}\right] \frac{%
\partial \Phi _{1}}{\partial \xi }.  \label{16}
\end{equation}%
The above equations represent the components of the electric field drift.

Now, the next higher order, $\epsilon ^{5/2}$, leads to the following set of
equations:%
\begin{equation}
\frac{\partial n_{i1}}{\partial \tau }+l_{z}\frac{\partial (n_{i1}v_{iz1})}{%
\partial \xi }=\lambda \frac{\partial n_{i2}}{\partial \xi }-l_{y}\frac{%
\partial v_{iy2}}{\partial \xi }-l_{z}\frac{\partial v_{iz2}}{\partial \xi }%
=f_{1},  \label{17}
\end{equation}%
\begin{equation}
\frac{\partial v_{iz1}}{\partial \tau }+l_{z}v_{iz1}\frac{\partial v_{iz1}}{%
\partial \xi }=\lambda \frac{\partial v_{iz2}}{\partial \xi }-2\theta
_{i}l_{z}\frac{\partial n_{i2}}{\partial \xi }-l_{z}\frac{\partial \Phi _{2}%
}{\partial \xi }=f_{2},  \label{18}
\end{equation}

\begin{equation}
\frac{\partial n_{n1}}{\partial \tau }+l_{z}\frac{\partial (n_{n1}v_{nz1})}{%
\partial \xi }=\lambda \frac{\partial n_{n2}}{\partial \xi }-l_{y}\frac{%
\partial v_{ny2}}{\partial \xi }-l_{z}\frac{\partial v_{nz2}}{\partial \xi }%
=f_{3},  \label{19}
\end{equation}%
\begin{equation}
\frac{\partial v_{nz1}}{\partial \tau }+l_{z}v_{nz1}\frac{\partial v_{nz1}}{%
\partial \xi }=\lambda \frac{\partial v_{nz2}}{\partial \xi }+\alpha
_{pn}l_{z}\frac{\partial \Phi _{2}}{\partial \xi }-2\alpha _{pn}\theta
_{n}l_{z}\frac{\partial n_{n2}}{\partial \xi }=f_{4},  \label{20}
\end{equation}

\begin{equation}
v_{ix2}=v_{nx2}=0,  \label{23}
\end{equation}%
\begin{equation}
v_{iy2}=\frac{\lambda l_{y}}{\omega _{i}^{2}}\left[ 1+\frac{2\theta
_{i}l_{z}^{2}}{\lambda _{mp}^{2}}\right] \frac{\partial ^{2}\Phi _{1}}{%
\partial \xi ^{2}},  \label{24}
\end{equation}%
and 
\begin{equation}
v_{ny2}=-\frac{\lambda \alpha _{pn}l_{y}}{\omega _{n}^{2}}\left[ 1+\frac{%
2\alpha _{pn}\theta _{n}l_{z}^{2}}{\lambda _{mn}^{2}}\right] \frac{\partial
^{2}\Phi _{1}}{\partial \xi ^{2}}.  \label{25}
\end{equation}%
Eliminating the second-order perturbed quantities except $\Phi_2$ from the Eqs.~(16)-(19)
 and then utilizing the values of $n_{i1}$, $n_{n1}$ and $n_{e1}$,
we obtain: 
\begin{equation}
\frac{\partial n_{i2}}{\partial \xi }=\frac{2\lambda l_{z}^{2}}{\lambda
_{mp}^{4}}\frac{\partial \Phi _{1}}{\partial \tau }+\frac{3\lambda
^{2}l_{z}^{4}}{\lambda _{mp}^{6}}\Phi _{1}\frac{\partial \Phi _{1}}{\partial
\xi }+\frac{\lambda ^{2}l_{y}^{2}}{\lambda _{mp}^{2}\sigma _{i}^{2}}\left \{
1+\frac{2\theta _{i}l_{z}^{2}}{\lambda _{mp}^{2}}\right \} \frac{\partial
^{3}\Phi _{1}}{\partial \xi ^{3}}+\frac{l_{z}^{2}}{\lambda _{mp}^{2}}\frac{%
\partial \Phi _{2}}{\partial \xi },  \label{26}
\end{equation}%
\begin{align}
\frac{\partial n_{n2}}{\partial \xi }& =-\frac{2\lambda \alpha _{pn}l_{z}^{2}%
}{\lambda _{mn}^{4}}\frac{\partial \Phi _{1}}{\partial \tau }+\frac{3\alpha
_{pn}^{2}\lambda ^{2}l_{z}^{4}}{\lambda _{mn}^{6}}\Phi _{1}\frac{\partial
\Phi _{1}}{\partial \xi }  \notag \\
& -\frac{\alpha _{pn}l_{z}^{2}}{\lambda _{mn}^{2}}\frac{\partial \Phi _{2}}{%
\partial \xi }-\frac{\lambda ^{2}\alpha l_{y}^{2}}{\sigma _{n}^{2}\lambda
_{mn}^{2}}\left \{ 1+\frac{2\alpha _{pn}\theta _{n}l_{z}^{2}}{\lambda
_{mn}^{2}}\right \} \frac{\partial ^{3}\Phi _{1}}{\partial \xi ^{3}},
\label{27}
\end{align}%

\begin{equation}
\frac{\partial n_{e2}}{\partial \xi }=
\Phi _{1}\frac{\partial \Phi _{1}}{\partial \xi }+\frac{\partial \Phi _{2}}{\partial \xi }.  \label{28}
\end{equation}%

The next order of the Poisson equation is $\sim\epsilon^{2}$, and in stretched coordinates this yields,%
\begin{equation}
\frac{\partial ^{2}\Phi _{1}}{\partial \xi ^{2}}%
=N_{e0}n_{e2}+N_{n0}n_{n2}-n_{i2}.  \label{29}
\end{equation}%
Note that we assume the electrons to follow the Boltzmann density distribution,
\begin{equation}
n_e \simeq n_{e0} e^{\Phi} \simeq [1+ \Phi + \frac{1}{2} \Phi^2 +...],
\end{equation}
which yields
\begin{equation}
n_e \simeq [1+\epsilon \Phi_1 + \epsilon^2 (\frac{1}{2} \Phi_1^2 + \Phi_2)+...].
\end{equation}
Operating $\partial _{\xi }$ on the Poisson equation and then using Eqs.~(\ref%
{26})-(\ref{28}) along with Eq.~(28), we get the following nonlinear partial differential
equation in stretched coordinates $(\xi, \tau)$, 
\begin{equation}
P\frac{\partial \Phi _{1}}{\partial \tau }+Q\Phi _{1}\frac{\partial \Phi _{1}%
}{\partial \xi }+R\frac{\partial ^{3}\Phi _{1}}{\partial \xi ^{3}}=0,
\label{30}
\end{equation}
where
\begin{equation}
P=\lambda l_z^2 \left[ \frac{1}{\lambda _{mp}^{4}} +\frac{N_{n0}\alpha _{pn}}{\lambda _{mn}^{4}}%
\right], \notag	
\end{equation}
\begin{equation}
	Q = \frac{l_z^4}{2}\left[ (\frac{3\lambda ^{2}}{\lambda _{mp}^{6}} -
		\frac{2\theta _{i}l_{z}^{2}}{\lambda _{mp}^{6}}) -N_{e0} - N_{n0}(\frac{%
			3 \alpha _{pn}^{2} \lambda^2}{\lambda _{mn}^{6}}-
		\frac{2 \theta_{- }\alpha _{pn}^{2}l_{z}^{2}}{\lambda _{mn}^{6}})\right],  \notag	
\end{equation}
and
\begin{equation}
	R = \frac{1}{2} \left[ 1+\frac{\lambda ^{2}l_{y}^{2}}{%
		\lambda _{mp}^{2} \omega_i^{2}}\left \{ 1+\frac{2\theta
		_{i}l_{z}^{2}}{\lambda _{mp}^{2}}\right \}+\frac{N_{n0}\lambda ^{2}\alpha _{pn}l_{y}^{2}}{\omega
		_{n}^{2}\lambda _{mn}^{2}}\left \{ 1+\frac{\alpha _{pn}2\theta
		_{n}l_{z}^{2}}{\lambda _{mn}^{2}}\right \}  \right].  \notag	
\end{equation}
Equation~(29) can be expressed in a simpler form, viz.\ 
\begin{equation}
\frac{\partial \Phi _{1}}{\partial \tau }+A\Phi _{1}\frac{\partial \Phi _{1}%
}{\partial \xi }+B\frac{\partial ^{3}\Phi _{1}}{\partial \xi ^{3}}=0,
\label{32}
\end{equation}%
where $A=Q/P$, and $B=R/P.$
Equation~(\ref{32}) represents the KdV equation which describes the
evolution of weakly nonlinear obliquely propagating electrostatic perturbations
in magnetized NPIE plasma. The stationary solution of Eq.~(\ref{32}) is obtained by using the transformation, 
\begin{equation}
\eta =\xi -M_{0}\tau,  \label{33}
\end{equation}%
where $M_0$ is the constant normalized speed of a solitary wave in the moving frame $\eta$. Note that if the speed of the solitary structure is denoted by $U$ in the frame of the wave phase speed $c_{si}$, in the normalized form we obtain the Mach number $1<M_0=\frac{c_{si}+U}{c_{si}}$.
Then, the partial differential equation in stretched coordinates (30) becomes an ordinary
differential equation in the moving coordinate $\eta$,
\begin{equation}
-M_{0}\frac{d \Phi }{d \eta }+A\Phi \frac{d \Phi }{%
d \eta }+B\frac{d^{3}\Phi }{d \eta ^{3}}=0,  \label{34}
\end{equation}%
where $\Phi _{1}$ is replaced by $\Phi $ for
convenience. Equation~(32) admits the following single pulse soliton
solution,
\begin{equation}
\Phi =\Phi _{m}\sec h^{2}(\frac{\eta }{W}),  \label{35}
\end{equation}%
where $\Phi _{m}=3M_{0}/A$ and $W=\sqrt{4B/M_{0}}$ are the
amplitude and width of the nonlinear structure, respectively. The maximum
amplitude $\Phi_m$ must be smaller than one ($1$), because the KdV equation has been obtained by
using the small amplitude approximation. It is important to mention that the Mach number $M_{0},$ width $W$ and Maximum
amplitude $\Phi _{m}$ of the solitons are linked to each other by the
following relation: $W^{2}\Phi _{m}=12B/A.$

%%%%%%%%%%%%%%%%%%%%%%%%%%%%%%%%%%%%%%%%%%%%%%%%%%%%%%%%%%%%%%%%%%%%%%%
\section{Comments on IAWs in magnetized plasma}
%%%%%%%%%%%%%%%%%%%%%%%%%%%%%%%%%%%%%%%%%%%%%%%%%%%%%%%%%%%%%%%%%%%%%%%
If negative ions are ignored using $n_{n0}=0$ in Eq.~(30), it becomes the same as Eq.~(12) of Ref.~(6) when dissipation is not considered by putting the coefficient $C=0$. In this case, we have
\begin{equation}
\lambda=l_z.
\end{equation}
Note that $\lambda$ is denoted by $v_0$ in the Ref.~\cite{Shahmansouri2013}.
The coefficients $A$ and $B$ of Eq.~(30) reduce to,
\begin{equation}
	A=\frac{3l_{z}^{2}}{2\lambda }-\frac{\lambda ^{3}}{2l_{z}^{2}}=l_z,  \label{40}
\end{equation}%
and 
\begin{equation}
	B=\frac{\lambda ^{3}}{2l_{z}^{2}}[1+\frac{1-l_{z}^{2}}{\sigma _{i}}]=\frac{\lambda}{2},
	\label{41}
\end{equation}%
which are the same expressions as those mentioned in Ref.~\cite{Shahmansouri2013}.
The solution of Eq.~(30) yields the maximum normalized amplitude of the solitary pulse $1<\Phi_m =\frac{3 M_0}{A}$ because $1<M_0$ and $A=l_z<1$ in this case.  

An interesting point is that in the reductive perturbation technique, the
polarization drift effect in the dispersion relation does not appear in the lowest order 
in case of magnetized electron ion plasma. If we use the
Fourier transformation, the linear dispersion relation for IAWs in usual
unmagnetized electron ion plasma under quasi-neutrality, has the following form:
\begin{equation}
\omega ^{2}=\frac{k^{2}c_{si}^{2}}{1+ k^2 \lambda_{De}^2},  \label{38}
\end{equation}%
where $\lambda_{De}$ is the electron Debye shielding length.
The physical quantities are assumed to be proportional to $%
e^{i(zk_{z}-\omega t)}$ and $k_z=k$ in this case. We know that under the plane wave assumption, the set of linearized equations of magnetized electron ion plasma yields the linear
dispersion relation of IAWs under quasi-neutrality $(n_i\simeq n_e)$ in the following form:
\begin{equation}
\omega ^{2}=\frac{k^{2}c_{si}^{2}}{1+k_{\perp }^{2}\rho _{s}^{2}},  \label{39}
\end{equation}%
where $\rho _{s}^{2}=c_{si}^{2}/\Omega _{i}^{2}$, $\Omega_i=\frac{eB_0}{m_i c}$ and $k_{\perp }^{2}$ is the
perpendicular component of the wave vector. Polarization drift produces
a dispersion effect through the term $k_{\perp }^{2}\rho _{s}^{2}$. In RPM we
use an ordering such that the dispersion balances the nonlinearity to give
rise to solitons and dispersion becomes a higher-order effect. 
Therefore, in the lowest order (linear case), the
polarization drift effect does not appear. Using Fourier transformation and ignoring polarization
drifts, the linear dispersion relation for IAWs in NPIE plasma can be written for $k_{\perp}=0$
as, 
\begin{equation}
N_{e0}-\frac{N_{n0}k_{z}^{2}c_{sn}^{2}}{\omega ^{2}-k_{z}^{2}v_{tn}^{2}}-%
\frac{k_{z}^{2}c_{si}^{2}}{\omega ^{2}-k_{z}^{2}v_{ti}^{2}}=0,  \label{42}
\end{equation}%
which can be expressed as a fourth-order polynomial in $\omega $ as 
\begin{equation}
N_{e0}\omega
^{4}-\{N_{e0}v_{tn}^{2}+N_{e0}v_{ti}^{2}+N_{n0}c_{sn}^{2}+c_{si}^{2}%
\}k_{z}^{2}\omega
^{2}+k_{z}^{4}%
\{N_{n0}c_{sn}^{2}v_{ti}^{2}+N_{e0}v_{tn}^{2}v_{ti}^{2}+c_{si}^{2}v_{tn}^{2}%
\}=0,  \label{43}
\end{equation}%
where $v_{tj}^{2}=\sqrt{T_{j}/m_{j}}$ and $c_{sj}^{2}=\sqrt{T_{e}/m_{j}}$
are the thermal speeds corresponding to the $j$th ion
species. The dispersion relation Eq.~(\ref{43}) does not contain any effects of the ambient magnetic field.
{In Ref. \cite{Ichiki2001}, multi ion component plasmas were produced introducing two kind of negative ions in the electron ion plasmas of different species and the three ion acoustic wave modes were observed in each case. In the second experiment \cite{Ichiki2002}, the one-negative ion plasmas were produced and xenon-fluorene-electron $Xe^{+}-F^{-}-e$ plasma was discussed in detail. Main finding was that the three ion acoustic modes appear in the presence of two negative ion species and two ion acoustic modes appear in the plasma with one-negative ion species. The frequencies of linear modes were observed and their characteristics were discussed. On the other hand, we have presented theoretical derivation of the dispersion relation for the coupled two ion acoustic modes in magnetized one-negative ion plasma to show that how the effects of magnetic field disappear under the framework of RPM in the lowest order.}
%%%%%%%%%%%%%%%%%%%%%%%%%%%%%%%%%%%%%%%%%%%%%%%%%%%%%%%%%%%%%%%%%%%%%%%
\section{Numerical Results}
%%%%%%%%%%%%%%%%%%%%%%%%%%%%%%%%%%%%%%%%%%%%%%%%%%%%%%%%%%%%%%%%%%%%%%%
Application of the theoretical calculations to a suitable system of NPIE plasma can elaborate the physical picture more clearly. For this purpose, we consider the data of the experiment in which negative ions of fluorene $F^{-}$ were introduced in xenon $Xe^{+}$ plasma \cite{Ichiki2001, Ichiki2002}. In these experiments, we have 
$m_i=(131)m_p$, $m_n=(19)m_p$ where $m_p=1.67 \times 10^{-24}\;$g is the proton mass, and the ranges of parameters are $n_{e0} \simeq (8\times 10^{8}-2\times 10^{9})\;$cm$^{-3}$, $T_e \simeq (0.2-0.4)\;$eV, $T_n \simeq T_i$, and $T_i\simeq (0.1) T_e$. For illustration, we assume $B_{0}=3\times 10^{3}\;$Gauss   and  choose $n_{i0}=10^{9}\;$cm$^{_{-3}}$, $T_{i}=(0.1)T_e\;$eV, $T_{n}=T_i\;$eV, and $T_{e}=(0.3)\;$eV. Since $m_n< m_i$ in this plasma, therefore we have $c_{si}< c_{sn}$. The plasmas produced in laboratory experiments [45,46] were unmagnetized. However, here we want to analyze the RPM in a magnetized plasma. 
Therefore, the magnitude of the external magnetic field has been taken from another experiment with negative ions \cite{Kim2007} .

The linear dispersion relation Eq.~(\ref{43}) has been obtained using the Fourier transformation and the two ion acoustic modes are the normal modes of this system; one corresponding to positive ions and the other corresponding to negative ions. The frequencies $\omega$ ($rad/s$) are plotted versus the $z$-component of wave vector $k_{z}(cm^{-1})$ in Fig.~(1), using relation (\ref{43}) for different ratios of densities keeping the temperatures fixed. Since the electron temperature is larger than the ions temperature, Landau damping can be ignored within the fluid theory framework. In Fig.~(2), the frequencies $\omega$ ($rad/s$) are plotted versus  $k_{z}(cm^{-1})$ for different temperature ratios of positive and negative ions keeping the densities fixed. The larger frequencies in Figs.~(1) and (2) represent the ion acoustic mode corresponding to negative ions.

The soliton profiles are plotted in Figs.~(3), and (4) using Eq.~(33) for different density and temperature ratios of the positive and negative ion species. We choose one out of the four roots for the phase speed $\lambda$ of the acoustic mode by solving Eq.~(13) to determine the coefficients of the KdV equation and consider the normalized Mach number $M_0=(1.3)$, assuming that the soliton speed is non-zero in the frame of phase speed $\lambda$. Note that the values of $\lambda$ vary with the temperatures and densities. The value of the nonlinear coefficient $A$ turns out to be larger than one, and consequently, the maximum amplitude becomes smaller than one (1) in agreement with the initial small amplitude approximation. Figure (6) has been plotted using the same values of physical parameters which are given in caption of Fig.~(3) accept the value of Mach number. In Fig.~(6), we use $M_0=(1.1)$ and it shows that the soliton's dip decreases corresponding to smaller value of Mach number $M_0=(1.1)$. 

The
variation in obliqueness also modifies the amplitude and width of soliton as shown
in Fig.~($5$) for two different values of $l_{z}=0.96$ and $l_{z}=0.98$. The values of $\lambda$ determined from Eq.~(13) also changes with $l_z$. 

\begin{figure}[H]
    \centering
    \includegraphics{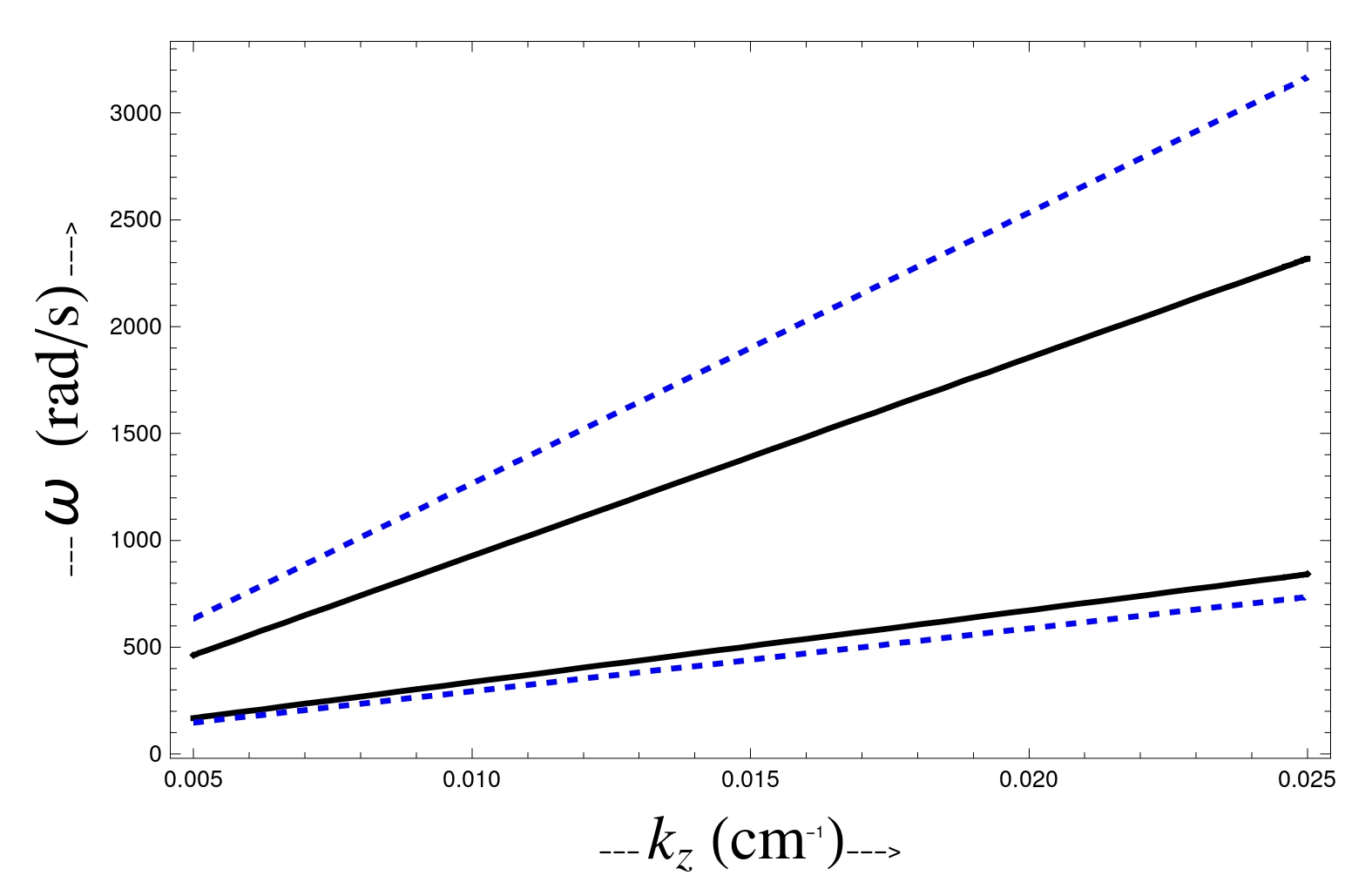}
    \caption{Frequencies $\omega $ vs $k_{z}(cm^{-1})$ are
plotted for $Xe^{+}-F^{-}-e$ plasma \textbf{(a)} $n_{e0}=(0.8) n_{i0}$ (solid black), 
\textbf{(b)} $%
n_{e0}=(0.6) n_{i0}$ (dotted blue) for $T_{e}=(0.3)\;$eV, $T_{i}=(0.1)T_e$, $T_n=2 T_i$, $n_{i0}=10^{9}\;$cm$^{-3}$ and $B_0= 10^{3}\;$G.}
    \label{fig:1}
\end{figure}

\begin{figure}[H]
    \centering
    \includegraphics{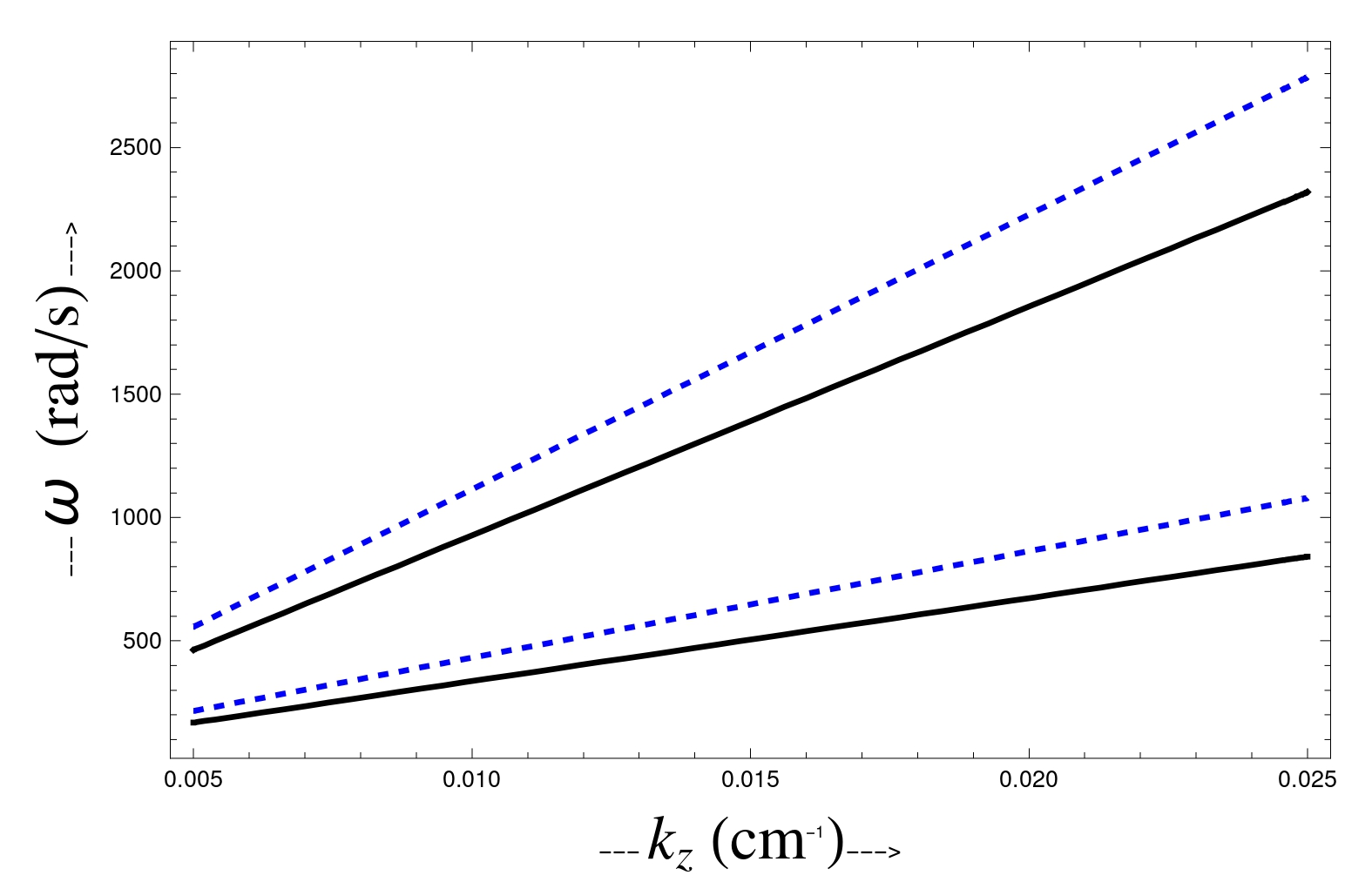}
    \caption{ Frequencies $\omega $ vs $k_{z}(cm^{-1})$ are
plotted for $Xe^{+}-F^{-}-e$ plasma \textbf{(a)} $T_{n}=2 T_{i}$ (solid black), \textbf{(b)} $T_{n}=5T_{i}$
(dotted blue) with $n_{e0}=(0.8) n_{i0}$ and other parameters are the same as in Fig.~(1).}
    \label{fig:2}
\end{figure}

\begin{figure}[H]
    \centering
    \includegraphics{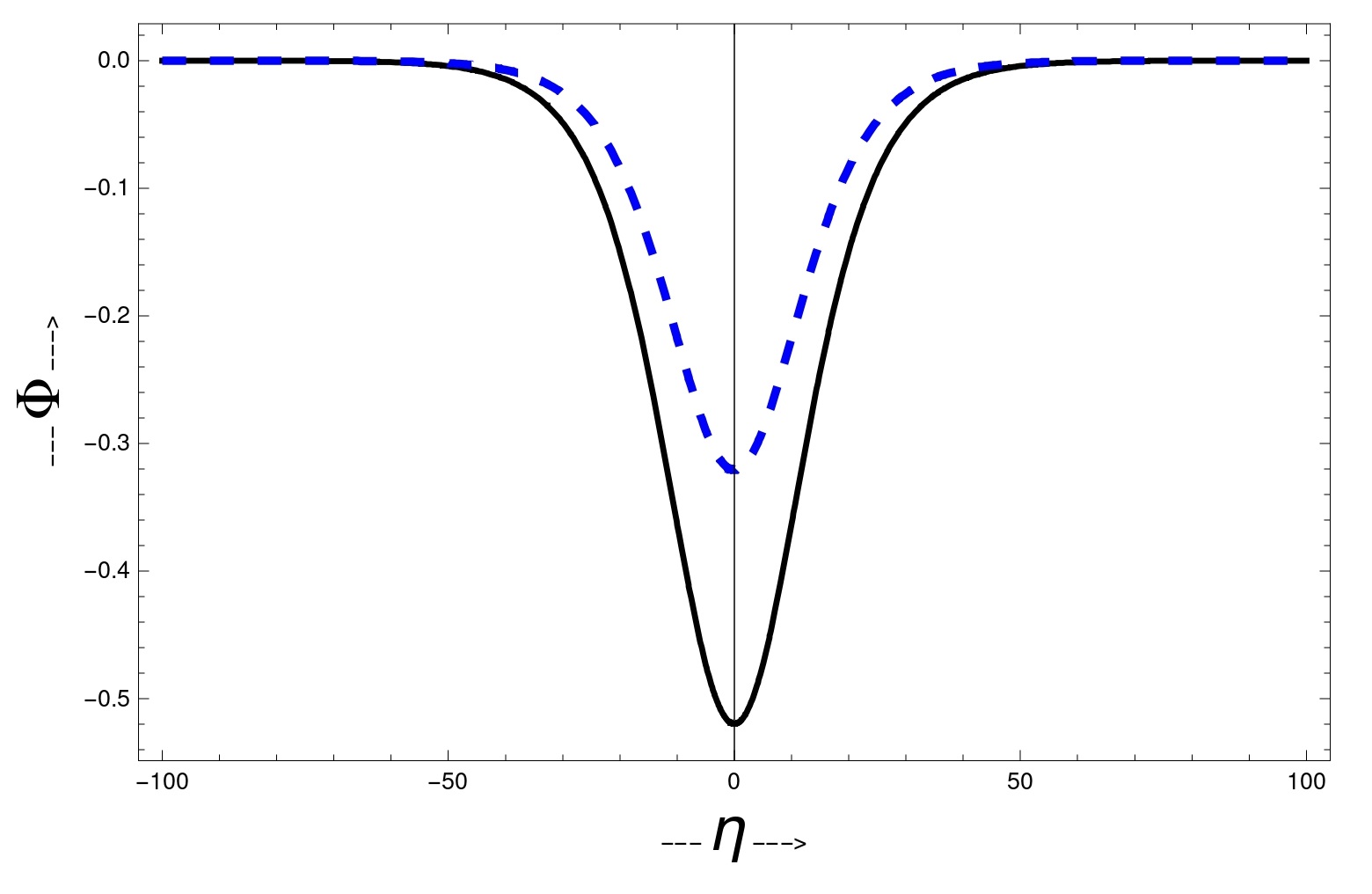}
    \caption{Soliton profile plotted corresponding to $T_{e}=(10) T_{i}=(0.3)\;$eV and
\textbf{(a)}  $T_n=2T_i$ (solid black), $\lambda =$ $2.2$, $l_{z}=0.98$,
and $n_{e0}=(0.8) n_{i0}$ \textbf{(b)} $T_{n}=5T_{i}$ (dashed blue), $\lambda
=$ $2.92$, $l_{z}=0.98$, $n_{e0}=(0.8) n_{i0}, $ and $M_{0}=1.3$. } 
    \label{fig:3}
\end{figure}

\begin{figure}[H]
    \centering
    \includegraphics{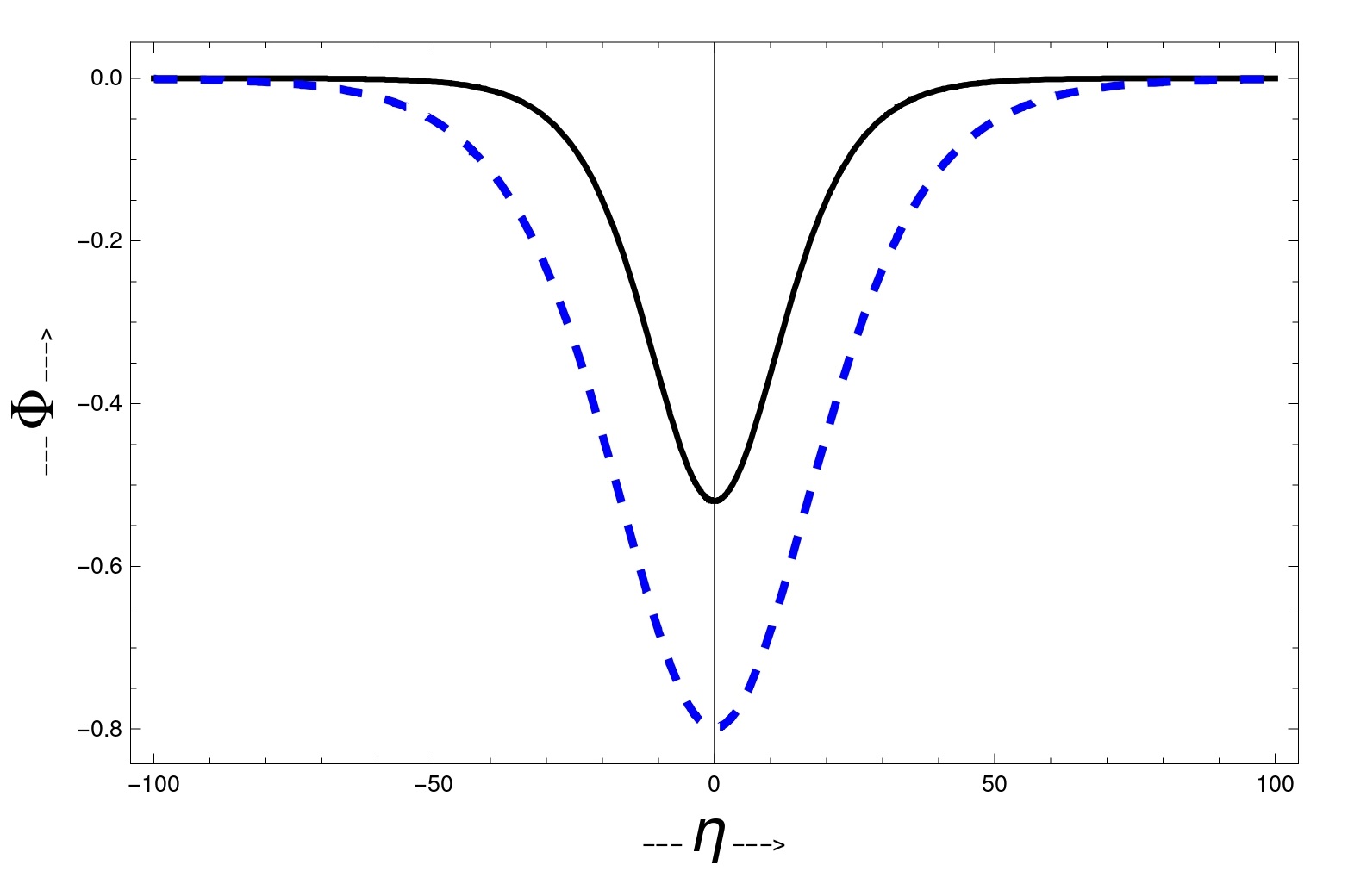}
    \caption{Soliton profile plotted corresponding to $T_{e}=(10) T_{i}=(0.3)\;$eV and
\textbf{(a)} $n_{e0}=(0.8) n_{i0}$ (solid black), $\lambda =$ $2.2$, $%
l_{z}=0.98$, and $T_{i}=(0.1) T_{e}$ \textbf{(b)} $n_{e0}=0.6 n_{i0}$ (dashed
blue), $\lambda =$ $2.85$, $l_{z}=0.98$, $ $ and $M_{0}=1.3$.} 
    \label{fig:4}
\end{figure}

\begin{figure}[H]
    \centering
    \includegraphics{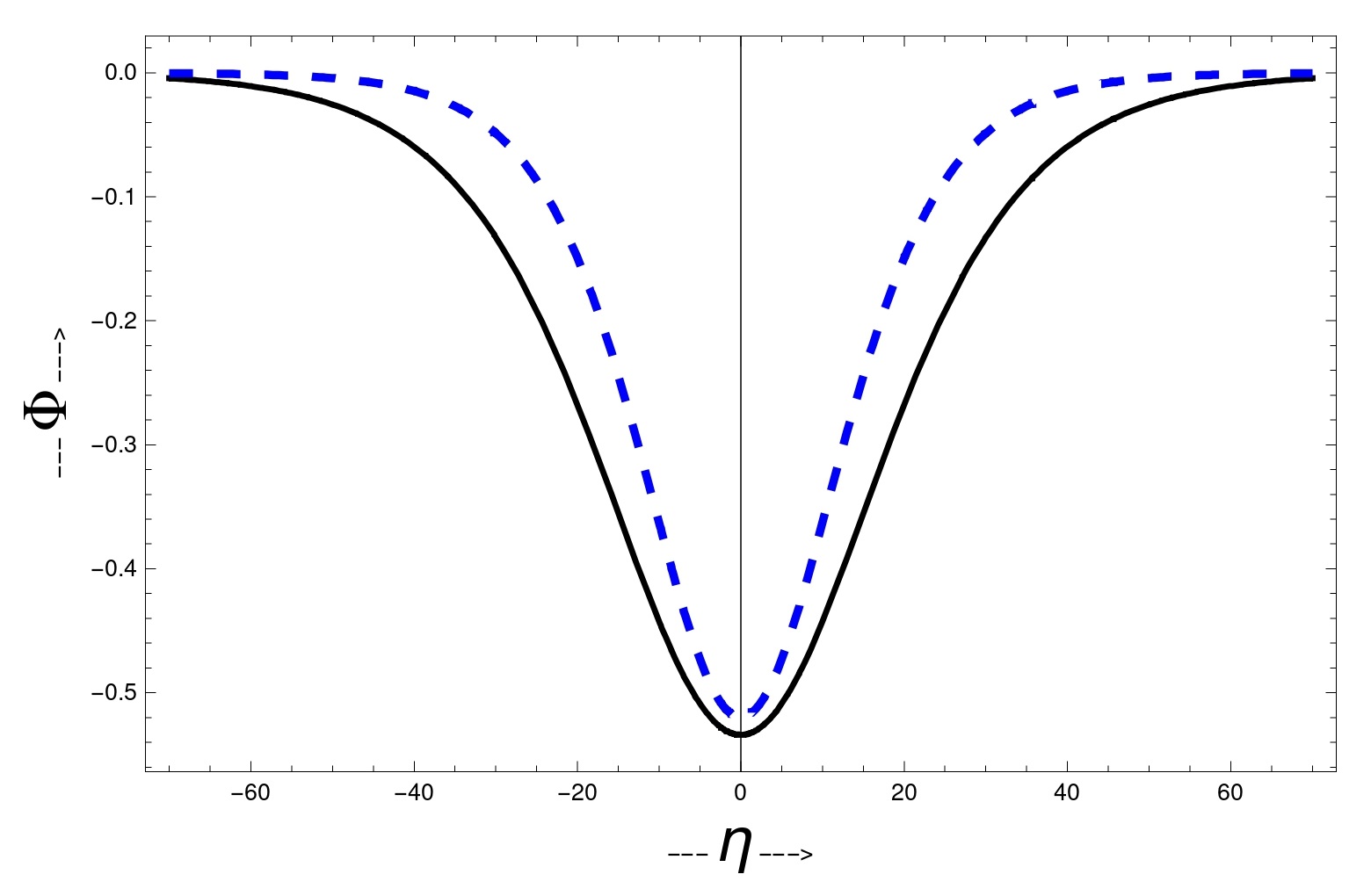}
    \caption{Soliton profile plotted corresponding to\
different values of obliqueness \textbf{(a)} $l_{z}=0.96$ (solid black), $%
\lambda =$ $2.16$, $T_{i}=(0.1) T_{e}$, and $n_{e0}=0.8n_{i0}$ \textbf{(b)} $%
l_{z}=0.98$ (dashed blue), $\lambda =$ $2.2$, $T_{i}=(0.1) T_{e}$, $%
n_{e0}=0.8n_{i0}, $ and  $M_{0}=1.3$. } 
    \label{fig:5}
\end{figure}

\begin{figure}[H]
    \centering
    \includegraphics{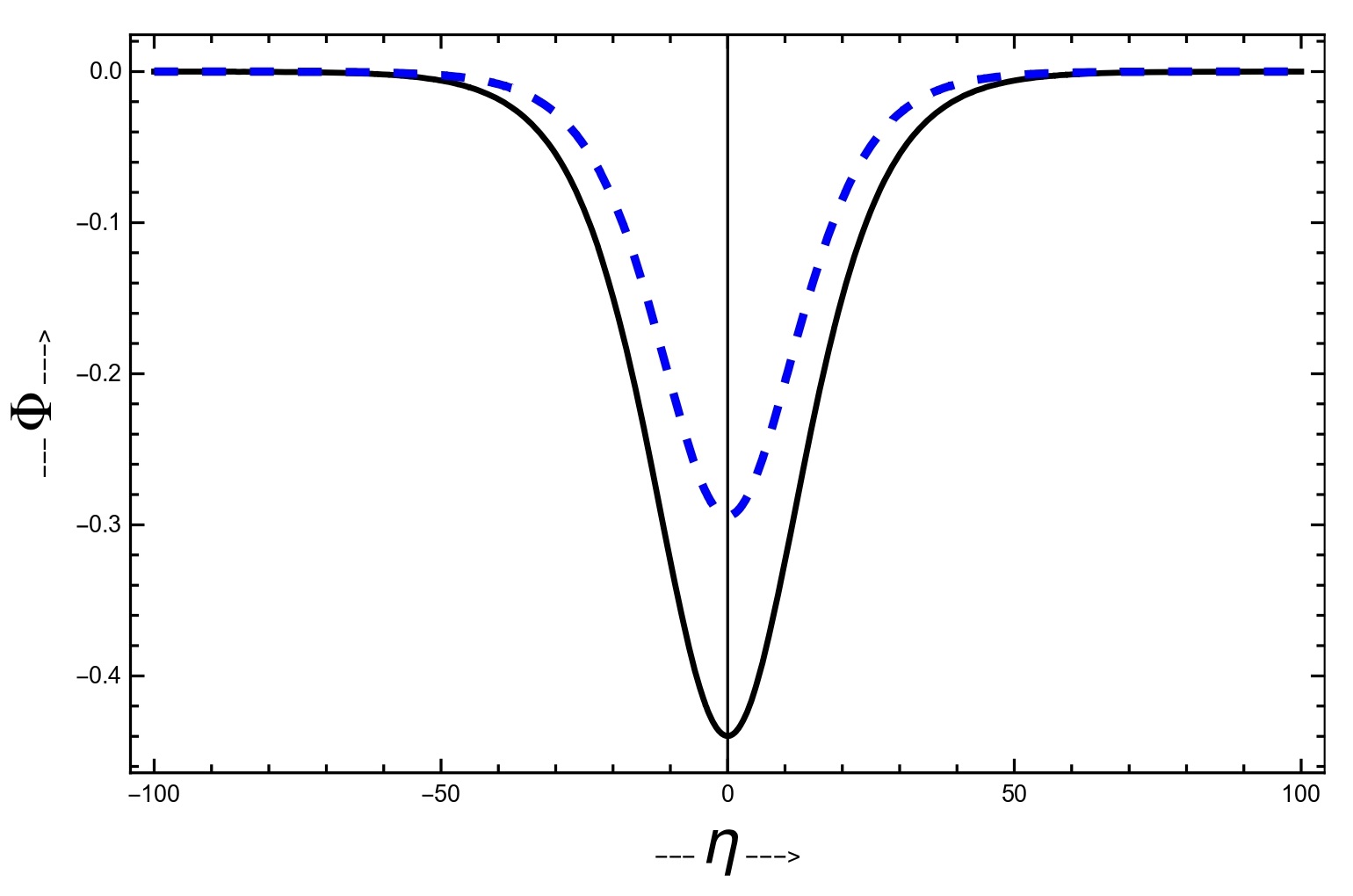}
    \caption{Soliton profile plotted corresponding to $T_{e}=(10) T_{i}=(0.3)\;$eV and
\textbf{(a)}  $T_n=2T_i$ (solid black), $\lambda =$ $2.2$, $l_{z}=0.98$,
and $n_{e0}=(0.8) n_{i0}$ \textbf{(b)} $T_{n}=5T_{i}$ (dashed blue), $\lambda
=$ $2.92$, $l_{z}=0.98$, $n_{e0}=(0.8) n_{i0}, $ and $M_{0}=1.1$.} 
    \label{fig:6}
\end{figure}

%\begin{table}[tbp] \centering%
	%EndExpansion
	%\caption{Amplitudes and Widths of solitons in Figs.~(4) %and (5)}%
	%\begin{tabular}{|l|l|l|l|l|l|}
		%\hline
		%Sr. No. & $n_{e0}/n_{i0}$ & $l_{z}$ & Phase Speed & %$\left \vert \Phi
		%_{m}\right \vert $ & $W(cm)$ \\ \hline
		%Fig.~4($a$) & $0.4$ & $0.98$ & $3.53$ & $0.79$ & %$15.13$ \\ \hline
		%Fig.~4($b$) & $0.7$ & $0.98$ & $2.09$ & $0.82$ & %$8.75$ \\ \hline
		%Fig.~5($a$) & $0.4$ & $0.96$ & $3.46$ & $0.91$ & %$20.65$ \\ \hline
		%Fig.~5($b$) & $0.4$ & $0.98$ & $3.53$ & $0.82$ & %$15.13$ \\ \hline
	%\end{tabular}%
	%\label{Parameters}%
	%TCIMACRO{\TeXButton{E}{\end{table}}}%
%BeginExpansion
%\end{table}%

\subsection{Role of negative ions} 
In the
presence of negative ions, the lowest order terms in Eq.~(13) yield a fourth order
polynomial in phase velocity $\lambda $ for IAWs given by the following equation:%
\begin{eqnarray}
&&\lambda ^{4}-\frac{n_{i0}}{n_{e0}}\{(2\theta _{i}+\alpha
_{pn}2\theta _{n})\frac{n_{i0}}{n_{e0}}l_{z}^{2}+(1+\frac{n_{n0}}{%
n_{i0}}\alpha _{pn})l_{z}^{2}\lambda ^{2}\}  \label{44} \\
&&\left. +\frac{n_{i0}}{n_{e0}}\{ \frac{n_{e0}}{n_{i0}}\alpha _{pn}4\theta _{n}\theta _{i}+\frac{n_{-0}}{n_{i0}}\alpha
_{pn}2\theta _{i}+\alpha _{pn}2\theta
_{n}\}l_{z}^{4}=0.\right.  \notag
\end{eqnarray}
To have the soliton structure given by Eq.~(33), we estimate the phase velocity $\lambda$ using the linear dispersion relation (13) obtained under RPM in the lowest order. The values of $\lambda$ represent two modes; one corresponding to xenon $Xe^{+}$ and one to $SF_6^{-}$ ions. Each mode has two branches, therefore we obtain 
$\lambda=\pm \lambda_1$ and $\lambda=\pm \lambda_2$. Since there are two IAW modes, the formation of solitary structures becomeś possible under the small amplitude limit if one of the modes has normalized phase speed larger than one. We choose $\lambda=\lambda_2>1$ to obtain solitary structures because $\lambda_1<1$.

For example in Fig.~(3), we find $\lambda_1=0.91$ and $\lambda_2=2.2$. Then, corresponding to $\lambda_2$, we get $A=-(7.5)$ which yields $\mid \Phi_m \mid =\frac{3M_0}{\mid A \mid} <1$ for $M_0=1.3$ and $\lambda_2=2.2$. On the other hand, for $\lambda_1=0.91$, the factor $\lambda_{mn}^2$ becomes negative and yields an unphysical result with imaginary value of $A$.

%%%%%%%%%%%%%%%%%%%%%%%%%%%%%%%%%%%%%%%%%%%%%%%%%%%%%%%%%%%%%%%%%%%%%%%
\section{Discussion}
%%%%%%%%%%%%%%%%%%%%%%%%%%%%%%%%%%%%%%%%%%%%%%%%%%%%%%%%%%%%%%%%%%%%%%%
The reductive perturbation method (RPM) has been analyzed by deriving the Korteweg-de Vries (KdV) equation for the nonlinear ion acoustic waves (IAWs) in magnetized ${\bf B}_0 \neq 0$ negative positive ion electron (NPIE) plasma. The case of usual electron ion plasma has also been discussed in the limit $n_{n0}=0$.
 It has been explained in detail in sections~4 and 5 that the KdV equation derived for IAWs in magnetized electron ion plasma suffers from an inconsistency in the framework of RPM because the maximum normalized amplitude of the nonlinear pulse turns out to be larger than one $1< \mid \Phi_m \mid$.
The solitary structures of nonlinear electrostatic ion acoustic waves (IAWs) have been investigated in magnetized negative positive ion electron (NPIE) plasma by deriving the Korteweg-de Vries (KdV) equation using the reductive perturbation method (RPM). Linear dispersion relations of IAWs have been compared using Fourier transformation and  the lowest order equations obtained by expanding physical parameters under the RPM approach. The NPIE plasma has two IAW modes; one corresponding to a larger acoustic speed and the other corresponding to a smaller acoustic speed. Both modes have two branches relative to their propagation along positive or negative direction with respect to the ambient magnetic field. It is well known that IAWs propagate making an angle with the external magnetic field in magnetized plasma.

It seems important to mention here that under the framework of the RPM, the KdV equation obtained for nonlinear ion acoustic waves does not admit the solitary structure solution with consistent physical assumptions in the simple case of a magnetized electron ion plasma with inertial ions and inertia-less electrons. As has been mentioned in section~4 and subsection~5.1, for such a plasma, the maximum normalized amplitude of the soliton becomes larger than one ($1<\mid \Phi_m \mid$) because the nonlinear coefficient turns out to be smaller than one ($\mid A \mid<1$). A few authors obtained solitary solutions of nonlinear IAWs in a magnetized electron ion plasma using the perturbation method, but they assumed the electrons to be non-thermal \cite{Sultana2010}. 

In the presence of negative ions in the plasma, there appear two ion acoustic modes and usually for one of the modes the normalized phase velocity becomes larger than one, so that the amplitude remains less than one. 
In NPIE plasma, Eq.~(13) yields two values of $\lambda$. For example, corresponding to the densities and temperatures used in Fig.~(3), we obtain $\lambda_1= 0.91$ and $\lambda_2=2.2$. Thus, for $\lambda_2$ and $M_0=1.3$, we obtain $l_z<A$, which yields $\mid \Phi_m \mid <1$, in agreement with the small amplitude approximation.
In Refs.~[45,46], the produced plasmas were unmagnetized and the main focus of the authors was to observe the slow and fast ion acoustic modes in the presence of negative ions in the plasma. They observed these modes by varying the ratios of densities and temperatures of different species. We have considered only one case, namely a $Xe^{+}-F^{-}$ plasma along with electrons [45]. The densities and temperatures mentioned in this experiment have been used to get numerical results by applying our theoretical model. However, the ambient magnetic field ${\bf B}_0$ has been assumed to be non-zero to analyze the RPM method in detail. It should also be noted that in magnetized electron ion plasma we have $\lambda_{De}<\rho_s$. But the dispersive term in KdV appears through the non-quasi-neutrality condition $\nabla \cdot {\bf E} \neq 0$ in KdV under the framework of RPM.

{It may be mentioned here that in one of the experiments on NPI plasmas \cite{Kim2007}, the electron attachement to the molecules of $C_7 F_{14}$ and 
$ S F_6$ was investigated in a thermally ionized potassium plasma. In another experiment \cite{Kim2008}, the elctrostatic ion cyclotron waves (ICWs) were observed and analyzed in a plasma containing positive ions of potassium $(K^{+})$, negative ions of perfluoromethylcyclohexane $(C_7 F_{14}^{-})$ and electrons. The ion cyclotron wave (ICW) appears in the magnetized ${\bf B}_0 \neq 0$ plasma and propagate predominenetly in the perpendicular direction with respect to ambient magnetic field. Later on , an experiment \cite{Kim2013} was performed to excite low frequency electrostatic waves in NPI plasma having $(K^{+})$,  $(C_7 F_{14}^{-})$ and electrons. It was pointed out that the characteristics of the excited electrostatic wave were neither similar to the ICW nor to the IAW. It was recognized by the experimenters as the electrostatic drift wave in a nearly electron-free plasma. Theoretical analysis of the drift wave in pure negative positive ion (NPI) plasma was also presented using kinetic approach \cite{Ros2013}. It is necessary to mention here that electrons contribution to the generation of drift waves cannot be ignored \cite{Che2013}, in general. The experiments mentioned above on NPI plasmas were focused only on the study of linear waves whereas we investigate the nonlinear ion acoustic waves in magnetized NPIE plasma to highlight the limitations of RPM. Furthermore, we have normalized the velocities in the set of equations with $c_{si}$ to get the normalized phase velocity $\lambda=1$ in the limit $n_{n}=0$ and $B_0=0$ which is mentioned in the literature \cite{Washimi1966} for nonlinear IAW under the framework of RPM. Therefore, to study nonlinear IAWs in magnetized plasma, we have considered the experimentally produced $(K^{+}-C_7 F_{14}^{-}-e)$ plasma where $m_n < m_{i}$ and added a magnetic field of magnitude $B_0 = 3 \times 10^{3}$ $G$ which was used in another NPI experimental plasma \cite{Kim2008}. }

This investigation will be useful for further experimental and theoretical work on NPIE and PIE plasmas. The comments on the RPM can also be helpful for further studies of small amplitude waves.

%%%%%%%%%%%%%%%%%%%%%%%%%%%%%%%%%%%%%%%%%%%%%%%%%%%%%%%%%%%%%%%%%%%%%%%
\section{Data Availability Statement}
%%%%%%%%%%%%%%%%%%%%%%%%%%%%%%%%%%%%%%%%%%%%%%%%%%%%%%%%%%%%%%%%%%%%%%%

%\begin{acknowledgments}
The data used for the preparation of the presented results has been taken from Refs.\ 45 and 46. 
%\end{acknowledgments}

\end{document}